\documentclass[12pt,ifthen]{svmult}

\usepackage{amsfonts}
\usepackage{tabularx}
\usepackage{graphics}

\usepackage{multicol}
\usepackage{mathtools}
\usepackage{bbm}

\usepackage{amssymb}
\usepackage{pdfrender}

\usepackage[super]{nth}

\usepackage[dvips]{epsfig}

\newcommand{\Ii}{\mathbbm{1}}

\usepackage{graphicx}
\usepackage{colortbl}
\usepackage{graphics}
\usepackage{animate}
\usepackage{color}
\usepackage{xcolor}
\usepackage{verbatim}
\usepackage{slashbox}
\usepackage{multirow}
\usepackage{lmodern}

\usepackage{float}

\newcommand\MyColorBoxY[1]{%
\setlength\fboxsep{1pt}
\colorbox{yellow}{#1}
}

\newcommand\MyColorBoxO[1]{%
\setlength\fboxsep{1pt}
\colorbox{red!50}{#1}
}

\newcommand\MyColorBoxB[1]{%
\setlength\fboxsep{1pt}
\colorbox{blue!20}{#1}
}

\newcommand\MyColorBoxG[1]{%
\setlength\fboxsep{1pt}
\colorbox{green!20}{#1}
}

\graphicspath{
{"G:/My Drive/CoursesColumbiaUniversity/ACPM/figures/"}
{"G:/My Drive/Personal/Conferences/FOUR/"}
}

\begin{document}

\title*{Post trade allocation: how much are bunched orders costing your performance?}
\titlerunning{Post trade allocation: how much are bunched orders costing your performance?}

%

\author{Ali Hirsa\inst{\thanks{Industrial Engineering and Operations Research, Columbia University, ali.hirsa@columbia.edu
Chief Scientific Officer, Ask2.ai, \texttt{ali.hirsa@ask2.ai}}} \and Massoud Heidari\inst{\thanks{Massoud Heidari contributed to this paper in his personal capacity.  The information, views, and opinions expressed herein are solely his own and do not necessarily represent the views of Point72, L.P. or its affiliates.  Point72, L.P. and its affiliates are not responsible for, and did not verify for accuracy, any of the information contained herein.}}}
\institute{}

\authorrunning{Ali Hirsa \& Massoud Heidari}

\maketitle

\vspace{-0.5in}

\begin{abstract}
Individual trade orders are often bunched into a block order for processing efficiency, where in post execution, they are allocated into individual accounts.  Since Regulators have not mandated any specific post trade allocation practice or methodology, entities try to rigorously follow internal policies and procedures to meet the minimum Regulatory ask of being procedurally fair and equitable.  However, as many have found over the years, there is no simple solution for post trade allocation between accounts that results in a uniform distribution of returns. Furthermore, in many instances, the divergences between returns do not dissipate with more transactions, and tend to increase in some cases. This paper is the first systematic treatment of trade allocation risk.  We shed light on the reasons for return divergence among accounts, and we present a solution that supports uniform allocation of return irrespective of number of accounts and trade sizes. 

\noindent
\textbf{Key words:} trade allocation; bunched orders; block trades, separately managed accounts; under/over allocation; optimal rounding; machine learning; fair and equitable; direct indexing; broker; investment manager; advisor; \texttt{CFTC}; \texttt{NFA}.

\end{abstract}



\section{Overview}
Often times, investors have a choice to invest in a co-mingled vehicle (\texttt{Fund}), or to invest in a separately managed account\footnote{\texttt{SMAs} are portfolios of individual securities managed by an asset management firm (see \cite{sma1}, \cite{sma2} for a more detailed description of \texttt{SMAs}).} (\texttt{SMA}) that mimics the performance of the underlying \texttt{Fund}. All investors in the \texttt{Fund}, regardless of their size and the time of their investment, receive the same future gross returns, without any tracking errors. The disadvantage is that the investors generally have to adhere to the terms and conditions of the \texttt{Fund}, as a limited partner. 

To retain control of their assets, transparency, and to be able to construct personalized portfolios\footnote{\texttt{SMAs} can be built specifically for each investor based on their personal investment goals and expectations, including the exclusion of any specified securities. Many asset management firms offer customized portfolios through \texttt{SMAs}. Here is a statement on Blackrock website, www.blackrock.com/us/financial-professionals/investment-strategies/managed-accounts, regarding their \texttt{SMAs}: ``{\em Blackrock Separately Managed Accounts (\texttt{SMAs}) provide enhanced capabilities to meet your clients' financial goals.}''}, more and more institutional clients have opted to invest in an \texttt{SMA}, while the manager directs investments through a sub-management contract. In most cases, \texttt{SMAs} and the \texttt{Fund} have the same investment objective, and the \texttt{SMAs} seek to replicate the returns of the \texttt{Fund}. 

In a \texttt{Fund}, all trades are tracked and accounted for at the \texttt{Fund} level, and gross returns of these trades are distributed among all clients in an identical manner.  In contrast, when dealing with \texttt{SMAs} all trades need to be allocated between \texttt{SMAs} in a proportional manner first, and the returns are calculated for each individual \texttt{SMA} separately. Have in mind this is also the case for \texttt{block trades} \cite{blocktrades}, as the \texttt{Fund} manager still needs to allocate those block trades into all accounts (in this case to the Fund also). 

In principle, this is a straightforward problem: allocate trades proportional to relative account size, and then calculate returns.  However, in practice, since trades hardly take place at a single price, or at the same time, and because one cannot divide securities into random fractions, allocations need to resort to rounding, which, as we will show, is the source of divergence in returns among accounts (and between \texttt{SMAs} and the \texttt{Fund}). For \texttt{block trades}, we can imagine situations where the manager pre-designs trading sizes to avoid fractional allocation and rounding but as the number of accounts increases with different account sizes would make it very challenging. The rounding forces a choice between accounts to be over-allocated and accounts to be under-allocated, which leads to uneven distribution of returns.  Furthermore, repeated application of a mechanical decision rule for the over/under allocation of trades (bias) will often-times lead to larger and larger divergence of returns among accounts and between the \texttt{SMAs} and the \texttt{Fund}.  To prevent large tracking errors, and to provide all accounts with equitable and even distribution of returns, the industry has adopted various approaches for allocation of trades, generally processed, at the end of each trading day.  This is often a manual and time-consuming process - the art of trade allocation - and does procedurally tries to ensure fair and equitable treatment of all accounts. The general premise is that the fair and equitable rationale is met via rigorously applied procedures that do not exhibit outward biases towards any of the accounts. However, as we will note in the paper, using any of the prevalent methodologies results in return divergence between accounts and the premise that the accounts balance out over time maybe flawed.

Thus our objective in this paper is to shed light on the problem of trade being allocated across multiple accounts or other vehicles. In specific, despite the mathematical rigor of the underlying process, we will use some simple examples to show the source of the problem, which leads to the divergence of returns, and present a solution to remedy the problem.

\section{\texttt{NFA/CFTC} Compliance Rule}

Although there are no general methodologies for trade allocation, the \texttt{NFA} Compliance Rule 2-10/\texttt{CFTC} Regulation 1.35, {\em the allocation of bunched orders for multiple accounts} \cite{nfaRules}, sets out three principles that should guide the allocation procedures: (1) fairness, (2) objectivity, and (3) timeliness.  In the same document, they provide examples of procedures that satisfy these objectives. Here is an excerpt from the \texttt{NFA/CFTC} Compliance Rule:

{\small \em ``\texttt{NFA} Compliance Rule 2-10 adopts by reference \texttt{CFTC} Regulation 1.35, Among other things, this regulation \textcolor{blue}{requires that bunched orders be allocated in a fair and equitable} manner so that no account or group of accounts consistently receives \textcolor{blue}{favorable or unfavorable} treatment over time. The rule further provides that Eligible Account Managers bear the \textcolor{blue}{responsibility for the fair and equitable allocation of bunched orders}." 
CFTC Regulation 1.35(b)(5)}

\newpage

Core Principles and Responsibilities\footnote{https://www.nfa.futures.org/rulebook/rules.aspx?RuleID=9029\&Section=9} (source: \texttt{NFA}):
\begin{itemize}
\item{The first, which arises in all such orders, involves the question of how the total number of contracts should be allocated to the various accounts included in the bunched order}
\item{The second issue involves the allocation of split or partial fills}
\item{The same set of core principles govern the procedures to be used in handling both of these issues. Any procedure for the general allocation of trades or the allocation of split and partial fills must be:}
\begin{itemize}
\item{designed to meet the overriding regulatory objective that allocations are non-preferential and are fair and equitable over time, such that no account or group of accounts receive consistently favorable or unfavorable treatment}
\item{sufficiently objective and specific to permit independent verification of the fairness of the allocations over time and that the allocation methodology was followed for any particular bunched order; and}
\item{timely, in that the Eligible Account Manager must provide the allocation information to FCMs that execute or clear the trade as soon as practicable after the order is filled and, in any event, sufficiently before the end of the trading day to ensure that clearing records identify the ultimate customer for each trade}
\end{itemize}
\end{itemize}
In short, \texttt{NFA} \& \texttt{CFTC} expect eligible account managers as fiduciaries to implement fair and equitable trade allocation methods.

\subsection{\texttt{NFA} \!cited\! examples}

Here are \texttt{NFA} cited examples of Allocation Methodologies (source: \texttt{NFA})

\begin{itemize}
\item{Example \#1 - \textcolor{blue!70}{Rotation of Accounts}, Rotation of accounts on a regular cycle, usually daily or weekly, which receive the most favorable fills.} 
\vspace{0.2in}
\item{Example \#2 - \textcolor{blue!70}{Random Allocation}, Computer generated random order of accounts and allocate the best price to the first account on the list and the worst to the last.}
\vspace{0.2in}
\item{Example \#3 - \textcolor{blue!70}{Highest Prices to the Highest Account} (\texttt{HPHA}) Numbers, Some firms rank accounts in order of their account numbers and then allocate the highest fill prices to the accounts with the highest account numbers.} 
\vspace{0.2in}
\item{Example \#4 - \textcolor{blue!70}{Average Price} (\texttt{APS}), Calculate the average price for each bunched order and then assign the average price to each allocated contract. In the alternative, the program will allocate the actual fill prices among the accounts included in the order to approximate, as closely as possible, the average fill price.} 
\vspace{0.2in}
\item{Other - As observed to be fair and equitable by the Eligible Account Manager.  Example, Rounding \textcolor{blue!70}{(simple and/or adjusted/alterative)}}
\end{itemize}

\section{Sample Portfolio for Illustrative Purposes}

Throughout this paper, we are going to use the sample portfolio presented in Table \ref{table:portfolioExample} to illustrate the issues of allocating trades (buy of sell) into multiple accounts that would lead to uneven distribution of returns and address shortcomings and weaknesses of the cited examples for trade allocation.
What follows applies to bunched orders and block trades for any traded instrument including single name equity, futures, derivatives, etc in any market.

\begin{figure}[H]
\centering
\includegraphics[scale = 0.55]{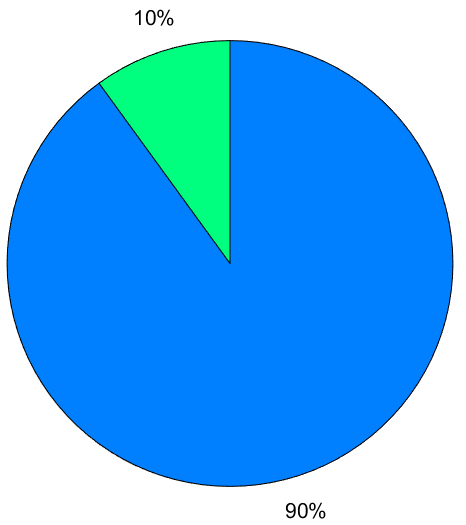}
\caption{Two accounts co-mingled in a fund}
\label{fig:Fund}
\end{figure}

\begin{figure}[H]
\centering
\includegraphics[scale = 0.55]{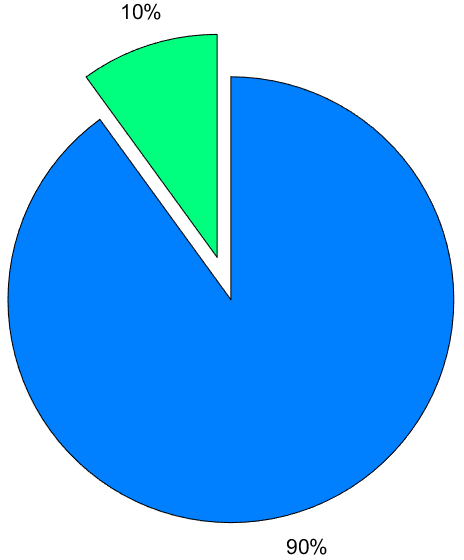}
\caption{Two separately managed accounts}
\label{fig:SMA}
\end{figure}

Table \ref{table:portfolioExample} represents trading of a single asset over two days in a fund. At the end of \nth{1} day, fund is short 4 with unrealized P\&L of \$380, and at the end of \nth{2} day, fund net position is zero (flat) with realized P\&L of \$1000.
\begin{table}[H]
\centering
\begin{tabular}{|c|c|c|c|c|c|} \hline
    & \textcolor{gray}{price}  & \textcolor{gray}{qty(B/S)} & \textcolor{gray}{net position}  & \textcolor{gray}{bucket P\&L}  & \textcolor{gray}{CumP\&L} \\
\raisebox{1.0ex}[0pt]{day} & $p_t$& $q_t$& $n\!p_t$ & $p\&l_t$ & $cp\&l_t$ \\ \hline\hline
\multirow{4}{*}{\rotatebox{0}{\nth{1} day}} & \$100   & 8   & 8        & -                                       & -   \\
&\$130   & 2   & 10       &  $8\!\times\!(\!130\!-\!100\!)\!=\!240$ & \$240 \\
&\$150   & -4  & 6        &  $10\!\times\!(\!150\!-\!130\!)\!=\!200$ & \$440 \\
&\$140   & -10 & -4       &   $6\!\times\!(\!140\!-\!150\!)\!=\!\mbox{-}60$ & \$380 \\  \hline\hline
\multirow{4}{*}{\rotatebox{0}{\nth{2} day}}&\$110   & -4 & -8       & $\mbox{-}4\!\times\!(\!110\!-\!140\!)\!=\!120$  & \$500 \\
&\$115   & -4  & -12      & $\mbox{-}8\!\times\!(\!115\!-\!110\!)\!=\!\mbox{-}40$  & \$460 \\
&\$110   & -4  & -16      & $\mbox{-}12\!\times\!(\!110\!-\!115\!)\!=\!120$  & \$580 \\
&\$80    & 16  & \MyColorBoxO{0}        & $\mbox{-}16\!\times\!(\!80\!-\!110\!)\!=\!480$ & \MyColorBoxY{\$1000} \\ \hline
\end{tabular}
\caption{Sample Portfolio}\label{table:portfolioExample}
\end{table}
For the fund in the sample portfolio, for simplicity, we assume it is comprised of two clients co-mingled in the fund as shown in Figure \ref{fig:Fund} with allocation factors\footnote{allocation factor for account $i$ for $i=1,\dots,N$ is calculated by 
$\alpha_{i} = \frac{\mbox{\sc{aum}}_{i}}{\sum_{i=1}^N \mbox{\sc{aum}}_{i}}$ where $\texttt{AUM}_i$ is asset under management (\texttt{AUM}) for account $i$} of 90\% \& 10\% respectively. 
Considering the two accounts are co-mingled in the fund, at the end of \nth{2} day, each account would receive P\&L according to its allocation factor, that is 
\begin{eqnarray*}
cumP\&L^{(1)} & = & \alpha_1\times  cumP\&L^{(\texttt{F})} = 0.90\times 1000 = \MyColorBoxY{\$900} \\
cumP\&L^{(2)} & = & \alpha_2\times  cumP\&L^{(\texttt{F})} = 0.10\times 1000 = \MyColorBoxY{\$100}
\end{eqnarray*}
Figure \ref{fig:SMA} shows the same clients who have opted for their individual separate accounts.  In a fund, all trades will be accounted for at the fund level, and clients will receive their proportional share of the P\&L, which leads to the identical return for all clients in the fund.  In 
contrast, trades need to be allocated between accounts before the 
calculation of return in each separate account.  In effect, clients of 
a fund have fractional virtual ownership of all transactions, whereas 
clients in an \texttt{SMA} need to be allocated rounded whole shares before 
calculation of return.

We should now allocate these bunched filled orders into these 2 accounts. With the preceding allocation factors, accounts' cumulative \texttt{P\&L} (\texttt{for an ideal scenario}) after allocation should ideally be (as before)
\begin{eqnarray*}
cumP\&L^{(1)} & = & 0.90\times 1000 = \MyColorBoxY{\$900} \\
cumP\&L^{(2)} & = & 0.10\times 1000 = \MyColorBoxY{\$100}
\end{eqnarray*}
Moreover, after the allocation on \nth{2} day, both accounts should have a net position of zero (flat) in order to be consistent with the fund. 

In our tests for assessment of allocation methodologies, we first start by \textcolor{blue}{simple rounding} followed by two sets of rounding with adjustment (non-trivial based on trial \& error). We then apply high price high account (\textcolor{blue}{\texttt{HPHA}}) and average system (\textcolor{blue}{\texttt{APS}}) for this sample portfolio. The last method that we present for testing is what we call ``{\em fair, optimal, and unbiased rounding}'' (\texttt{\textcolor{blue}{FOUR}}) that is our proposed machine learning (\texttt{ML}) algorithm for allocation of bunched orders into multiple accounts. Our assessment and conclusion are based on findings, observations, performance of these methodologies, and takeaways from these tests. We do not illustrate the test results of the Rotation of Accounts and Random Allocation methods as we believe the methodology names are sufficient in explaining their mathematical shortcomings and resultant performance divergence between accounts.

\subsection{Simple rounding for two accounts}

Our \nth{1} trial is a simple rounding. In Table \ref{table:simpleRounding}, we show the results of the simple rounding to this sample portfolio. We first multiply each filled order (buy or sell) by accounts' allocation factors and round them to a whole number. After rounding we have to make sure that summation is preserved\footnote{Assume hypothetically total buy filled orders is $5$ with allocation factors of 90\% and 10\%, which would imply $0.90\times 5 = 4.5$ and $0.10\times 5 = 0.5$. After a simple rounding, we get whole numbers $5$ and $1$ respectively but that would yield  $5+1 = 6 \neq 5$. In order to preserve total buy orders $5$ after rounding, we have to round up one and round down the other one, i.e. $5$\&$0$ or $4$\&$1$ which either way would be over-allocating one and under-allocating the other one.}.
\begin{table}[H]
\centering
\begin{tabular}{|>{\columncolor[gray]{0.8}}c|>{\columncolor[gray]{0.8}}c|>{\columncolor[gray]{0.8}}c|>{\columncolor[gray]{0.8}}c|>{\columncolor[gray]{0.8}}c||>{\columncolor[gray]{0.9}}c|>{\columncolor[gray]{0.9}}c|>{\columncolor[gray]{0.9}}c|>{\columncolor[gray]{0.9}}c||>{\columncolor[gray]{0.7}}c|>{\columncolor[gray]{0.7}}c|>{\columncolor[gray]{0.7}}c|>{\columncolor[gray]{0.7}}c|} \hline
\multicolumn{5}{|c||}{Fund} & \multicolumn{4}{c||}{1st account} & \multicolumn{4}{c|}{2nd account} \\ \hline \hline
$t$                                &$p_t$ &$q_t$&$n\!p_t$&$cp\&l_t$&$q_t^{(1)}$&$n\!p_t^{(1)}$&$p\&l_t^{(1)}$& $cp\&l_t^{(1)}$&$q_t^{(2)}$&$n\!p_t^{(2)}$&$p\&l_t^{(2)}$                                &$cp\&l_t^{(2)}$  \\ \hline\hline
                                   &\$100 & 8  & 8       & -       & 7.2(7)    & 7                 & -   & -                    &  0.8(1)  & 1  							& -  & -\\
                                   &\$130 & 2  & 10      & \$240   & 1.8(2)    & 9                 & 210 & \$210                &  0.2(0)  & 1  							& 30 &  \$30\\
\raisebox{1.25ex}[0pt]{\nth{1} day}&\$150 & -4 & 6       & \$440   &-3.6(-4)   & 5                 & 180 & \$390                & -0.4(0)  & 1  							& 20 &  \$50\\
                                   &\$140 & -10& -4      & \$380   & -9.0(-9)  & -4                & -50 & \$340                & -1.0(-1) & 0  							&-10 &  \$40\\ \hline\hline
                                   &\$110 & -4 & -8      & \$500   & -3.6(-4)  & -8                & 120 & \$460                & -0.4(0)  & 0  							& 0  &  \$40\\
                                   &\$115 & -4 & -12     & \$460   & -3.6(-4)  & -12               & -40 & \$420                & -0.4(0)  & 0  							& 0  &  \$40\\								   
\raisebox{1.25ex}[0pt]{\nth{2} day}&\$110 & -4 & -16     & \$520   & -3.4(-4)  & -16               &  60 & \$480                & -0.4(0)  & 0  							& 0  &  \$40\\
                                   &\$80  & 16 & 0       & \$1000  &  14.4(14) & \MyColorBoxO{-2}  & 480 & \MyColorBoxY{\$960} &  1.6(2)  & \MyColorBoxO{2} & 0  & \MyColorBoxY{\$40}\\ \hline
\end{tabular}
\caption{Simple Rounding for trade allocation into two accounts}\label{table:simpleRounding}
\end{table}
As we see in Table \ref{table:simpleRounding}, P\&Ls are not close to the ideal case (\$960, \$40 as opposed to \$900, \$100) and aside from discrepancy in accounts' P\&Ls, one account is short two versus the other one being long two even though the portfolio is flat.

\subsection{Rounding with adjustment for two accounts}

We now try a rounding with adjustment. Results of this rounding are depicted in Table \ref{table:manipulativeRounding}. Highlighted boxes show adjustment to rounding. For example, we see in this trial we round -3.6 to -3 as opposed to -4 and in order to preserve the summation that is -4, we are obligated to round -0.4 to -1 as opposed to 0. This is a non-trivial rounding. Just think of how challenging that would become as the number of accounts increases. One may wonder why just adjusting these two filled orders. We can definitely try a few other trials but as mentioned these tests are just for illustrative purposes to address the issue with rounding of filled orders.
\begin{table}[H]
\centering
\begin{tabular}{|>{\columncolor[gray]{0.8}}c|>{\columncolor[gray]{0.8}}c|>{\columncolor[gray]{0.8}}c|>{\columncolor[gray]{0.8}}c|>{\columncolor[gray]{0.8}}c||>{\columncolor[gray]{0.9}}c|>{\columncolor[gray]{0.9}}c|>{\columncolor[gray]{0.9}}c|>{\columncolor[gray]{0.9}}c||>{\columncolor[gray]{0.7}}c|>{\columncolor[gray]{0.7}}c|>{\columncolor[gray]{0.7}}c|>{\columncolor[gray]{0.7}}c|} \hline
\multicolumn{5}{|c||}{Fund} & \multicolumn{4}{c||}{1st account} & \multicolumn{4}{c|}{2nd account} \\ \hline \hline
$t$                                &$p_t$ &$q_t$&$n\!p_t$&$cp\&l_t$&$q_t^{(1)}$&$n\!p_t^{(1)}$&$p\&l_t^{(1)}$& $cp\&l_t^{(1)}$&$q_t^{(2)}$&$n\!p_t^{(2)}$&$p\&l_t^{(2)}$                                &$cp\&l_t^{(2)}$  \\ \hline\hline
                                   &\$100 & 8         & 8        & -       &7.2(7)    & 7                 & -   & -                   &  0.8(1)  & 1  							  & -  & -\\
                                   &\$130 & 2         &10        & \$240   & 1.8(2)    & 9                 & 210 & \$210               &  0.2(0)  & 1  							  & 30 & 30\\
\raisebox{1.25ex}[0pt]{\nth{1} day}&\$150 & -4        & 6        & \$440   &\MyColorBoxB{-3.6(\textcolor{red}{-3})}  & 6                 & 180 & \$390               & \MyColorBoxB{-0.4(\textcolor{red}{-1})} & 0  							  & 20 & 50\\
                                   &\$140 & -10       & -4       & \$380   &-9.0(-9)  & -3                & -60 & \$330               & -1.0(-1) & -1  							&  0 & 50\\ \hline\hline
                                   &\$110 & -4        & -8       & \$500   &-3.6(-4)  & -7                &  90 & \$420               & -0.4(0)  & -1  							& 30 & 80\\
                                   &\$115 & -4        & -12      & \$460   &-3.6(-4)  & -11               & -35 & \$385               & -0.4(0)  & -1  							& -15& 75\\
\raisebox{1.25ex}[0pt]{\nth{2} day}&\$110 & -4        & -16      & \$520   &\MyColorBoxB{-3.6(\textcolor{red}{-3})}  & -14               &  55 & \$440               & \MyColorBoxB{-0.4(\textcolor{red}{-1})} & -2  							& 5  & 80\\
                                   &\$80  & 16        & 0        & \$1000  &14.4(14)  & \MyColorBoxO{0}   & 420 & \MyColorBoxY{\$860} &  1.6(2)  & \MyColorBoxO{0} & 60 & \MyColorBoxY{\$140}\\
\hline
\end{tabular}
\caption{Adjusted rounding for allocation into two accounts}\label{table:manipulativeRounding}
\end{table}
The allocations for this adjusted rounding are much better as with this adjustment both accounts are flat now but P\&Ls are still off. It begs the question, can we  do better?

We try another adjustment to rounding which is more complicated than the previous one and non-trivial. In this rounding attempt, we adjust rounding in 5 of the filled orders (recall we adjusted just  2 in the previous example). Results from this trial are shown in Table \ref{table:manipulative2Rounding}.
\begin{table}[H]
\centering
\begin{tabular}{|>{\columncolor[gray]{0.8}}c|>{\columncolor[gray]{0.8}}c|>{\columncolor[gray]{0.8}}c|>{\columncolor[gray]{0.8}}c|>{\columncolor[gray]{0.8}}c||>{\columncolor[gray]{0.9}}c|>{\columncolor[gray]{0.9}}c|>{\columncolor[gray]{0.9}}c|>{\columncolor[gray]{0.9}}c||>{\columncolor[gray]{0.7}}c|>{\columncolor[gray]{0.7}}c|>{\columncolor[gray]{0.7}}c|>{\columncolor[gray]{0.7}}c|} \hline
\multicolumn{5}{|c||}{Fund} & \multicolumn{4}{c||}{1st account} & \multicolumn{4}{c|}{2nd account} \\ \hline \hline
$t$                        &$p_t$ &$q_t$&$n\!p_t$&$cp\&l_t$&$q_t^{(1)}$&$n\!p_t^{(1)}$&$p\&l_t^{(1)}$& $cp\&l_t^{(1)}$&$q_t^{(2)}$&$n\!p_t^{(2)}$&$p\&l_t^{(2)}$                                &$cp\&l_t^{(2)}$  \\ \hline\hline
                                   &\$100 & 8  & 8        & -       & \MyColorBoxB{7.2(\textcolor{red}{8})}    & 8                 & -   & -                   &  \MyColorBoxB{0.8(\textcolor{red}{0})}   & 0               & -  & -\\
                                   &\$130 & 2  &10        & \$240   & 1.8(2)                                   & 10                & 240 & \$240               &  0.2(0)                                 & 0               & 0  & 0\\
\raisebox{1.25ex}[0pt]{\nth{1} day}&\$150 & -4 & 6        & \$440   & \MyColorBoxB{-3.6(\textcolor{red}{-3})}  & 7                 & 200 & \$440               & \MyColorBoxB{-0.4(\textcolor{red}{-1})} & -1              & 0  & 0\\
                                   &\$140 & -10& -4       & \$380   &\MyColorBoxB{-9.0(\textcolor{red}{-10})} & -3                & -70 & \$370               & \MyColorBoxB{-1.0(\textcolor{red}{0})}  & -1              & 10 & 10\\ \hline\hline
                                   &\$110 & -4 & -8       & \$500   & -3.6(-4)                                 & -7                & 90  & \$460               & 0                                       & -1              & 30 & 40\\
                                   &\$115 & -4 & -12      & \$460   & -3.6(-4)                                 & -11               & -35 & \$425               & -0.4(0)                                 & -1              & -5 & 35\\
\raisebox{1.25ex}[0pt]{\nth{2} day}&\$110 & -4 & -16      & \$520   & \MyColorBoxB{-3.6(\textcolor{red}{-3})}  & -14               &  55 & \$480               & \MyColorBoxB{-0.4(\textcolor{red}{-1})} & -2 			  & 5  & 40\\
                                   &\$80  & 16 & 0        & \$1000  & 14.4(14)                                 & \MyColorBoxO{0}   & 420 & \MyColorBoxY{\$900} &  1.6(2)                                 & \MyColorBoxO{0} & 60 & \MyColorBoxY{\$100}\\
\hline
\end{tabular}
\caption{\nth{2} Adjustment to rounding for trade allocation into two accounts}\label{table:manipulative2Rounding}
\end{table}
As we see, this trial yields ideal and perfect results, but the rounding was non-trivial and fully based on trial \& error. It is very important to note that in reality one should not expect perfect and ideal outcomes.

\subsection{High Price High Account (\texttt{HPHA})}

High Price High Account (\texttt{HPHA}) is used for exchanges/markets that do not allow averaging for trades throughout the day. \texttt{HPHA} is used to allocate the highest price fills of the same order to the largest account and the lowest price fills to the smallest account.  In addition, industry participants use {\em rotation} and {\em randomization} of accounts to alleviate biases as a result of favorable fills which is allowed by regulators.

Accounts are ranked according to their account size and then the highest filled prices are allocated to the accounts with the highest allocation factors. Any advantage the higher numbered accounts enjoy on the sell order are presumably offset by the disadvantage on the buy orders. There is a note by \texttt{CFTC/NFA} stating that {\em ``under certain market conditions this may not always be true, but the method generally complies with the standards''}.

\subsubsection{Applying \texttt{HPHA} to the Sample Portfolio}

On \nth{1} day, we sort buy and sell orders as shown in Table \ref{table:sorted1}. 
\begin{table}[H]
\centering
\begin{tabular}{|>{\columncolor[gray]{0.8}}c|>{\columncolor[gray]{0.8}}c|} \hline
\multicolumn{2}{|c|}{Sorted Prices} \\
\multicolumn{2}{|c|}{(Buy)} \\ \hline
$p_t$&$q_t$\\ \hline \hline
130 & 2\\
100 & 8\\   \hline
\end{tabular}
\quad
\begin{tabular}{|>{\columncolor[gray]{0.8}}c|>{\columncolor[gray]{0.8}}c|} \hline
\multicolumn{2}{|c|}{Sorted Prices} \\
\multicolumn{2}{|c|}{(Sell)} \\ \hline
$p_t$&$q_t$\\ \hline \hline
150 & -4\\
140 & -10  \\ \hline
\end{tabular}
\caption{Sorted buy and sell orders on \nth{1} day}\label{table:sorted1}
\end{table}
Following the description of high price high account we allocate as follows:
\begin{itemize}
\item{For sell orders: one sell order at 140 gets allocated to \nth{2} account (smaller account) and the rest would get allocated to \nth{1} account (larger account) as shown in Table \ref{table:sorted1_allocated}.}
\item{For buy orders: one buy order at 100 gets allocated to \nth{2} account and the rest would get allocated to \nth{1} account as shown in Table \ref{table:sorted1_allocated}.}
\end{itemize}
These allocations are highlighted in Table \ref{table:sorted1_allocated} (the only difference between Tables \ref{table:sorted1} \& \ref{table:sorted1_allocated} is the highlighted allocation for buy and sell orders).

\begin{table}[H]
\centering
\begin{tabular}{|c|>{\columncolor[gray]{0.8}}c|>{\columncolor[gray]{0.8}}c|} \hline
\multicolumn{3}{|c|}{} \\
\multicolumn{3}{|c|}{\raisebox{1.25ex}[0pt]{Sorted Prices (Buy)}} \\ \hline
 Account &  $p_t$&$q_t$\\ \hline \hline
         &   130 & 2\\
\raisebox{1.25ex}[0pt]{\nth{1} acct} &   100 & 7\\ \hline\hline
\raisebox{0.0ex}[0pt]{\nth{2} acct}  &   100 & \MyColorBoxY{\textcolor{red}{1}} \\ \hline
\end{tabular}
\quad
\begin{tabular}{|c|>{\columncolor[gray]{0.8}}c|>{\columncolor[gray]{0.8}}c|} \hline
\multicolumn{3}{|c|}{} \\
\multicolumn{3}{|c|}{\raisebox{1.25ex}[0pt]{Sorted Prices (Sell)}} \\ \hline
 Account  & ~~$p_t$~~&$q_t$\\ \hline \hline
          & 150      & -4\\
\raisebox{1.25ex}[0pt]{\nth{1} acct} & 140 &  -9 \\ \hline\hline
\raisebox{0.0ex}[0pt] {\nth{2} acct} &140 & \MyColorBoxY{\textcolor{red}{-1}}  \\ \hline
\end{tabular}
\caption{Allocation of buy and sell orders on \nth{1} day}\label{table:sorted1_allocated}
\end{table}

On \nth{2} day, we do the same, sort buy and sell orders as shown in Table \ref{table:sorted2}.

\begin{table}[H]
\centering
\begin{tabular}{|>{\columncolor[gray]{0.8}}c|>{\columncolor[gray]{0.8}}c|} \hline
\multicolumn{2}{|c|}{Sorted Prices} \\
\multicolumn{2}{|c|}{(Buy)} \\ \hline
$p_t$&$q_t$\\ \hline \hline
                             &   \\ 
\raisebox{1.250ex}[0pt]{80}  & \raisebox{1.250ex}[0pt]{16}\\ \hline
\end{tabular}
\quad
\begin{tabular}{|>{\columncolor[gray]{0.8}}c|>{\columncolor[gray]{0.8}}c|} \hline
\multicolumn{2}{|c|}{Sorted Prices} \\
\multicolumn{2}{|c|}{(Sell)} \\ \hline
$p_t$&$q_t$\\ \hline \hline
115 & -4\\
110 & -8\\ \hline
\end{tabular}
\caption{Sorted buy and sell orders on \nth{2} day}\label{table:sorted2}
\end{table}

Following the description of high price high account we allocate as follows:
\begin{itemize}
\item{For sell orders: one sell order at 110 gets allocated to \nth{2} account (smaller account) and the rest would get allocated to \nth{1} account (larger account).}
\item{For buy orders: two buy orders at 80 gets allocated to \nth{2} account and the rest would get allocated to \nth{1} account}
\end{itemize}
These allocations are highlighted in Table \ref{table:sorted2_allocated} (the only difference between Tables \ref{table:sorted2} \& \ref{table:sorted2_allocated} is the highlighted allocation for buy and sell orders).
\begin{table}[H]
\centering
\begin{tabular}{|c|>{\columncolor[gray]{0.8}}c|>{\columncolor[gray]{0.8}}c|} \hline
\multicolumn{3}{|c|}{} \\
\multicolumn{3}{|c|}{\raisebox{1.25ex}[0pt]{Sorted Prices (Buy)}} \\ \hline
Account & $~~p_t$~~&$q_t$\\ \hline \hline
\raisebox{-1.250ex}[0pt]{\nth{1} acct}&  \raisebox{-1.250ex}[0pt]{80}  & \raisebox{-1.250ex}[0pt]{14}\\
                                      &      &   \\ \hline\hline
\raisebox{0.0ex}[0pt] {\nth{2} acct}  &  80  & \MyColorBoxY{\textcolor{red}{2}}\\ \hline
\end{tabular}
\quad\quad
\begin{tabular}{|c|>{\columncolor[gray]{0.8}}c|>{\columncolor[gray]{0.8}}c|} \hline
\multicolumn{3}{|c|}{} \\
\multicolumn{3}{|c|}{\raisebox{1.25ex}[0pt]{Sorted Prices (Sell)}} \\ \hline
Account & ~~$p_t$~~&$q_t$\\ \hline \hline
     & 115 & -4\\
\raisebox{1.25ex}[0pt] {\nth{1} acct}     & 110 & -7\\ \hline \hline
\raisebox{0.0ex}[0pt] {\nth{2} acct}      & 110 & \MyColorBoxY{\textcolor{red}{-1}}\\ 
\hline
\end{tabular}
\caption{Allocation of buy and sell orders on \nth{2} day}\label{table:sorted2_allocated}
\end{table}
Having allocations, we can calculate cumulative P\&L for each account
\begin{eqnarray*}
cumP\&L^{(1)} & = & (4\!\times\!150\!+\!9\!\times\!140\!+\!4\!\times\!115\!+\!7\!\times\!110) - (2\!\times\!130\!+\!7\!\times\!100\!+\!14\!\times\!80)= \MyColorBoxY{\$1010} \textcolor{gray!40}{~~\mbox{vs.} ~\$900}\\
cumP\&L^{(2)} & = & (0\!\times\!150\!+\!1\!\times\!140\!+\!0\!\times\!115\!+\!1\!\times\!110) - (0\!\times\!130\!+\!1\!\times\!100\!+\!2\!\times\!80) = \MyColorBoxY{-\$10}\textcolor{gray!40}{~~\mbox{vs.} ~\$100}
\end{eqnarray*}
Clearly the method is neither fair nor equitable and aside from discrepancy in P\&Ls, one account is short one contract versus the other one being long one even though the portfolio is flat.

To make it better, we do a slight adjustment (based on trial \& error) for allocating trades on \nth{1} day (highlighted in \MyColorBoxG{green}\hspace{-0.05in}) and keep the allocation intact on the \nth{2} day as shown in Tables \ref{table:sorted1_allocatedNew} and \ref{table:sorted2_allocatedNew}. Note that Tables \ref{table:sorted2_allocated} and \ref{table:sorted2_allocatedNew} are identical as we do not make any change to its allocation.

\begin{table}[H]
\centering
\begin{tabular}{|c|>{\columncolor[gray]{0.8}}c|>{\columncolor[gray]{0.8}}c|} \hline
\multicolumn{3}{|c|}{} \\
\multicolumn{3}{|c|}{\raisebox{1.25ex}[0pt]{Sorted Prices (Buy)}} \\ \hline
Account & ~~$p_t$~~&$q_t$\\ \hline \hline
        & 130 & 2\\
\raisebox{1.25ex}[0pt]{\nth{1} acct}        & 100 & 7\\ \hline\hline
\nth{2} acct        & 100 & \MyColorBoxY{\textcolor{red}{1}} \\ \hline
\end{tabular}
\quad
\begin{tabular}{|c|>{\columncolor[gray]{0.8}}c|>{\columncolor[gray]{0.8}}c|} \hline
\multicolumn{3}{|c|}{} \\
\multicolumn{3}{|c|}{\raisebox{1.25ex}[0pt]{Sorted Prices (Sell)}} \\ \hline
Account  & ~~$p_t$~~&$q_t$\\ \hline \hline
  &150   & -4\\
\raisebox{1.25ex}[0pt]{\nth{1} acct}  &140   & \MyColorBoxG{-8} \\ \hline\hline
\nth{2} acct  &140   & \MyColorBoxG{\textcolor{red}{-2}}  \\ \hline
\end{tabular}
\caption{Allocation of buy and sell orders on \nth{1} day}\label{table:sorted1_allocatedNew}
\end{table}

\begin{table}[H]
\centering
\begin{tabular}{|c|>{\columncolor[gray]{0.8}}c|>{\columncolor[gray]{0.8}}c|} \hline
\multicolumn{3}{|c|}{\raisebox{0.0ex}[0pt]{}} \\
\multicolumn{3}{|c|}{\raisebox{1.25ex}[0pt]{Sorted Prices (Buy)}} \\ \hline
Account &$p_t$&$q_t$\\ \hline \hline
     &     &     \\
\raisebox{1.25ex}[0pt]{\nth{1} acct}     & \raisebox{1.25ex}[0pt]{80}  & \raisebox{1.25ex}[0pt]{14}\\ \hline\hline
\raisebox{0.0ex}[0pt]{\nth{2} acct}     & 80  & \MyColorBoxY{\textcolor{red}{2}}\\ \hline
\end{tabular}
\quad\quad
\begin{tabular}{|c|>{\columncolor[gray]{0.8}}c|>{\columncolor[gray]{0.8}}c|} \hline
\multicolumn{3}{|c|}{Sorted Prices} \\
\multicolumn{3}{|c|}{(Sell)} \\ \hline
Account & $p_t$&$q_t$\\ \hline \hline
\raisebox{-1.25ex}[0pt]{\nth{1} acct}     &115 & -4\\
     &110 & -7\\  \hline\hline
\raisebox{0.0ex}[0pt]{\nth{2} acct}     &110 & \MyColorBoxY{\textcolor{red}{-1}}\\ 
\hline
\end{tabular}
\caption{Allocation of buy and sell orders on \nth{2} day}\label{table:sorted2_allocatedNew}
\end{table}
With this minor adjustment, we get the following cumulative P\&Ls for each account:
\begin{eqnarray*}
cumP\&L^{(1)} & = & (4\!\times\!150\!+\!\MyColorBoxG{8}\!\times\!140\!+\!4\!\times\!115\!+\!7\!\times\!110) - (2\!\times\!130\!+\!7\!\times\!100\!+\!14\!\times\!80)= \MyColorBoxY{\$870} \textcolor{gray!40}{~~\mbox{vs.} ~\$900}\\
cumP\&L^{(2)} & = & (0\!\times\!150\!+\!\MyColorBoxG{2}\!\times\!140\!+\!0\!\times\!115\!+\!1\!\times\!110) - (0\!\times\!130\!+\!1\!\times\!100\!+\!2\!\times\!80) = \MyColorBoxY{\$130}\textcolor{gray!40}{~~\mbox{vs.} ~\$100}
\end{eqnarray*}
This is not ideal but the outcome is much better and in this case both accounts are flat at the end of \nth{2} day trading consistent with the fund being flat.

\subsection{Average Pricing System (\texttt{APS})}
 
For exchanges (e.g. \texttt{CME}) and markets that allow averaging of price, an average price is used for all 
transactions within the same order.  In effect, the \texttt{APS} increases the trade size to reduce 
the bias, but in no way alleviate the need to round. However, as the previous section makes clear, the return variation in trade allocation is a result of biases due to rounding, and neither of these procedures address those concerns.

Average prices of buy and sell orders are calculated as follows per each trading day:
\begin{eqnarray*}
\overline{P}_B & = & \frac{\sum_{j=1}^J q_j\Ii_{\{q_j>0\}} \times p_{j}}{\sum_{j=1}^J q_j\Ii_{\{q_j>0\}}}  
                 = \frac{\sum_{j=1}^J q_j\Ii_{\{q_j>0\}} \times p_{j}}{Q_B} \\
\overline{P}_S & = & \frac{\sum_{j=1}^J q_j\Ii_{\{q_j<0\}} \times p_{j}}{\sum_{j=1}^J q_j\Ii_{\{q_j<0\}}}  
                 = \frac{\sum_{j=1}^J q_j\Ii_{\{q_j<0\}} \times p_{j}}{Q_S}
\end{eqnarray*}
where $p_j$ represents price and $q_j$ the quantity filled at that price, for long (buy order) $q_j$ is positive and for short (sell order) $q_j$ is positive. $Q_B$ and $Q_S$ are total buy and sell orders for the trading day respectively. As before $\alpha_{i}$ is the allocation factor for account $i$.


For $N$ managed accounts $i=1,\dots,N$, we should calculate
\begin{eqnarray*}
q^{(i)}_B & = & \alpha_{i}\times Q_B \\
q^{(i)}_S & = & \alpha_{i}\times Q_S 
\end{eqnarray*}
and then round $q^{(i)}_B$ and $q^{(i)}_S$ while preserving the sum, that is $Q_B$ and $Q_S$ respectively. At the first glance it seems that \texttt{APS} may resolve the issue on rounding. First question is what happens as the number of accounts increases? Discrepancy could easily come from the rounding step as the number of accounts increases and gets exaggerated for overnight positions.

\subsubsection{Applying \texttt{APS} to the Sample Portfolio}

On \nth{1} day: average buy \& sell prices for total \texttt{BUY} of $Q_B=10$ and total \texttt{SELL} of $Q_S=14$ are:
\begin{eqnarray*}
\mbox{average \texttt{BUY} price}  & = & \frac{8\!\times\!100+2\!\times\!130}{8+2} = \frac{1060}{10} = \textcolor{red}{106} \\
\mbox{average \texttt{SELL} price} & = & \frac{(\mbox{-}4)\!\times\!150 + (\mbox{-}10)\!\times\!140}{(\mbox{-}4) + (\mbox{-}10)} = \frac{2000}{14} = \textcolor{red}{142.8571}
\end{eqnarray*}
and round the quantities to get
\[
\begin{array}{lll}
\mbox{\nth{1} acct \texttt{BUY} qty}  \!=\!  0.90\!\times\!10 \!=\! 9.5(\textcolor{red}{9}) & \quad\quad & \mbox{\nth{1} acct \texttt{SELL} qty}  \!=\!  0.90\!\times\!14 \!=\! 12.6(\textcolor{red}{13}) \\
\mbox{\nth{2} acct \texttt{BUY} qty}  \!=\!  0.05\!\times\!10 \!=\! 0.5(\textcolor{red}{1}) & \quad\quad & \mbox{\nth{2} acct \texttt{SELL} qty}  \!=\!  0.10\!\times\!14 \!=\! 1.4(\textcolor{red}{1})
\end{array}
\]
On \nth{2} day: average buy \& sell prices for total \texttt{BUY} of $Q_B=16$ and total \texttt{SELL} of $Q_S=12$ are:
\begin{eqnarray*}
\mbox{average \texttt{BUY} price}  & = & \frac{16\!\times\!80}{16} = \frac{1280}{16} = \textcolor{red}{80} \\
\mbox{average \texttt{SELL} price} & = & \frac{(\mbox{-}8)\!\times\!110+(\mbox{-}4)\!\times\!115}{(\mbox{-}8)+(\mbox{-}4)} = \frac{1340}{12} = \textcolor{red}{111.6667}
\end{eqnarray*}
and round the quantities to get
\[
\begin{array}{lll}
\mbox{\nth{1} acct \texttt{BUY} qty}  \!=\!  0.90\!\times\!16 \!=\! 14.4(\textcolor{red}{14}) & \quad\quad &  \mbox{\nth{1} acct \texttt{SELL} qty}  \!=\!  0.90\!\times\!12 \!=\! 10.8(\textcolor{red}{11}) \\
\mbox{\nth{2} acct \texttt{BUY} qty}  \!=\!  0.10\!\times\!16 \!=\! 1.6(\textcolor{red}{2})   & \quad\quad &  \mbox{\nth{2} acct \texttt{SELL} qty}  \!=\!  0.10\!\times\!12 \!=\! 1.2(\textcolor{red}{1})
\end{array}
\]
Having average prices and quantities on both trading days, we can calculate cumulative P\&Ls for each managed account
\begin{eqnarray*}
cumP\&L^{(1)} & = & (13\!\times\!142.8571+11\!\times\! 111.6667)-(9\!\times\!106+14\!\times\!80)= \colorbox{yellow}{\$1011.48} \textcolor{gray!40}{~~\mbox{vs.} ~\$900}\\
cumP\&L^{(2)} & = & (1\!\times\!142.8571+1\!\times\!111.6667)-  (1\!\times\!106+2\!\times\!80) = \colorbox{yellow}{-\$11.48} \textcolor{gray!40}{~~\mbox{vs.} ~\$100}
\end{eqnarray*}
We see by simply following the procedure in \texttt{APS}, the allocation would not be fair and equitable. Aside from discrepancy in P\&Ls, one account is short one contract versus the other one being long one contract even though portfolio net position is zero.

Let's try \texttt{APS} with some non-trivial adjustments on the rounding procedure. The aim is to try to make allocation to be more fair and equitable (trial \& error).

\noindent On \nth{1} day: we do the following non-trivial rounding on quantities
\[
\begin{array}{lll}
\mbox{\nth{1} acct \texttt{BUY} qty}  \!=\!  0.90\!\times\!10 \!=\! 9.5(\colorbox{yellow}{\textcolor{red}{10}}) & \quad\quad & \mbox{\nth{1} acct \texttt{SELL} qty}  \!=\!  0.90\!\times\!14 \!=\! 12.6(\colorbox{yellow}{\textcolor{red}{12}}) \\
\mbox{\nth{2} acct \texttt{BUY} qty}  \!=\!  0.05\!\times\!10 \!=\! 0.5(\colorbox{yellow}{\textcolor{red}{0}}) & \quad\quad & \mbox{\nth{2} acct \texttt{SELL} qty}  \!=\!  0.10\!\times\!14 \!=\! 1.4(\colorbox{yellow}{\textcolor{red}{2}})
\end{array}
\]
On \nth{2} day: we do the following non-trivial rounding on quantities
\[
\begin{array}{lll}
\mbox{\nth{1} acct \texttt{BUY} qty}  \!=\!  0.90\!\times\!16 \!=\! 14.4(\textcolor{red}{14}) & \quad\quad &  \mbox{\nth{1} acct \texttt{SELL} qty}  \!=\!  0.90\!\times\!12 \!=\! 10.8(\colorbox{green}{\textcolor{red}{12}}) \\
\mbox{\nth{2} acct \texttt{BUY} qty}  \!=\!  0.10\!\times\!16 \!=\! 1.6(\textcolor{red}{2})   & \quad\quad &  \mbox{\nth{2} acct \texttt{SELL} qty}  \!=\!  0.10\!\times\!12 \!=\! 1.2(\colorbox{green}{\textcolor{red}{0}})
\end{array}
\]
The non-trivial rounding is highlighted in \colorbox{green}{green}. Having these quantities we can calculate cumulative P\&Ls for each account and compare them against the ideal scenario
\begin{eqnarray*}
cumP\&L^{(1)} & = & (12\!\times\!142.8571+12\!\times\! 111.6667)-(10\!\times\!106+14\!\times\!80) = \colorbox{yellow}{\$874.29} \textcolor{gray!40}{~~\mbox{vs.} ~\$900}\\
cumP\&L^{(2)} & = & ( 2\!\times\!142.8571+ 0\!\times\! 111.6667)-( 0\!\times\!106+ 2\!\times\!80) = \colorbox{yellow}{\$125.71} \textcolor{gray!40}{~~\mbox{vs.} ~\$100}
\end{eqnarray*}

With these non-trivial adjustments on rounding we get much better results, both accounts are flat and P\&Ls looks much better.

\section{\textcolor{blue}{\texttt{F}}air \textcolor{blue}{\texttt{O}}ptimal \textcolor{blue}{\texttt{U}}nbiased \textcolor{blue}{\texttt{R}}ounding (\texttt{FOUR})}\label{sec:four}

By going through this sample portfolio and utilizing different allocation methods we can see with no adjustment on rounding, none of the methods would have yielded fair and equitable results. It begs the question, can there be an optimal adjustment on rounding? Are there any golden rules for alteration on rounding?

This is what precisely has been addressed and developed in \cite{ourPatent}. A machine learning (\texttt{ML}) algorithm that by construction finds fair, optimal, and unbiased rounding (\texttt{FOUR}) in allocating every bunched filled order (buy or sell) or block orders. 

The task in trade allocation is to allocate each trade (buy or sell) such that proportional returns are distributed evenly and nearly identically among pari-passu accounts. The main idea in \texttt{FOUR} is the inclusion of the return-to-time-of-trade in the allocation procedure and done by tracking running P\&L for accounts in order to correct over/under allocation over time. 

Here are the steps in \texttt{FOUR}. For each block orders or bunched filled order (buy or sell), we first find an allocation vector based on a rounding routine\footnote{See \cite{optimalRounding} for optimal rounding under 
integer constraints.}. Recall that in rounding we should make sure to preserve the summation for that order. We then calculate the trade P\&Ls for the allocation vector and add  them to prior cumulative P\&Ls 
to update cumulative P\&Ls for each account up to that time, say $t$. Having the running P\&Ls for each account at $t$, we calculate the following quantity called $\mathbb{Q}_t$ by
\begin{eqnarray*}
{\mathbb Q}_t =  \sum_{i=1}^{N} \left( PnL_{t} - \fcolorbox{green}{yellow}{$\frac{PnL^{(i)}_{t}}{\alpha_{i}}$}\right)^2
\end{eqnarray*}
where
\begin{itemize}
\item[]{$PnL_{t}^{(i)}$: cumulative P\&L of account $i$ at time $t$}
\item[]{$PnL_{t}$: aggregated P\&L at time $t$ (fund P\&L)}
\item[]{$\alpha_i$: allocation factor for account $i$, for $i=1,\dots,N$}
\end{itemize}
Note that normalizing each account cumulative P\&L, $PnL_t^{(i)}$, by its allocation factor, $\alpha_i$, and comparing against aggregated cumulative P\&L, $PnL_{t}$, would ensure minimum variation of return across all accounts. This is also equivalent to writing $\mathbb{Q}_t$ as 
\begin{eqnarray*}
{\mathbb Q}_t =  \sum_{i=1}^{N} \left( r_{t} - r^{(i)}_{t} \right)^2
\end{eqnarray*}
where
\begin{itemize}
\item[]{$r^{(i)}_t$: return of account $i$ at time $t$}
\item[]{$r_t$: return of the fund at time $t$}
\end{itemize}
Adjustment to trade allocation is done through an \texttt{ML} algorithm in a narrow range. By doing this adjustment in a narrow range, we limit the search space which is user specified. Over/under allocation impact on P\&Ls might not get corrected by adjustment of just a single share/contract. Therefore, we may need to do a wider search which means perhaps penalizing one account more versus other accounts as long as $\mathbb{Q}_t$ is being reduced. Allocations are supposed to be either  non-negative or non-positive to assure accounts do not buy or sell from  each other. Therefore, a sanity check is needed after each adjustment to assure the obtained allocation vector is admissible. For each of those obtained allocation vectors, we update cumulative P\&Ls and re-evaluate $\mathbb{Q}_t$. Optimal allocation vector for the filled order at time $t$ is the one with the smallest value for $\mathbb{Q}_t$. This procedure in the \texttt{ML} algorithm gets repeated for each filled order until the very last trade for that day.

In case of very low frequency trading, we can apply the following reinforcement:
\begin{eqnarray*}
{\mathbb Q}_t =  \sum_{i=1}^{N} w_{t}^{(i)} \left( r_{t} - r^{(i)}_{t} \right)^2
\end{eqnarray*}
with
\begin{eqnarray*}
w_{t}^{(i)} =  \frac{\exp(\delta_i)}{\sum_{j=1}^{N}\exp(\delta_j)}
\end{eqnarray*}
where $\delta_i=|r_{t} - r^{(i)}_{t}|$. The hyperparameter $\delta_i$ can get updated weekly, monthly, or quarterly (user specified).

We see that the optimal allocation is obtained by keeping track of return variation across all accounts to ensure minimum variation of returns among them. It is important to notice as soon as each block order or bunched order gets filled, the algorithm in \texttt{FOUR} can be utilized to find the optimal allocation for that order which implies \texttt{FOUR} can be applied in real time. The procedure can be applied to each individual traded instrument, group of instruments, or at the fund level, depending on how accounts are set up. That would facilitate trade allocation into customized portfolios.

\subsubsection{Applying \texttt{FOUR} to the Sample Portfolio}

Table \ref{table:four} shows allocations utilizing \texttt{FOUR} by following the procedure described earlier for the method (we omit showing step-by-step procedure in trade allocation and just showing the final outcome of the \texttt{ML} algorithm).
\begin{table}[H]
\centering
\begin{tabular}{|>{\columncolor[gray]{0.8}}c|>{\columncolor[gray]{0.8}}c|>{\columncolor[gray]{0.8}}c|>{\columncolor[gray]{0.8}}c|>{\columncolor[gray]{0.8}}c||>{\columncolor[gray]{0.9}}c|>{\columncolor[gray]{0.9}}c|>{\columncolor[gray]{0.9}}c|>{\columncolor[gray]{0.9}}c||>{\columncolor[gray]{0.7}}c|>{\columncolor[gray]{0.7}}c|>{\columncolor[gray]{0.7}}c|>{\columncolor[gray]{0.7}}c|} \hline
\multicolumn{5}{|c||}{Fund} & \multicolumn{4}{c||}{1st account} & \multicolumn{4}{c|}{2nd account} \\ \hline \hline
$t$ & $p_t$&$q_t$&$n\!p_t$&$cp\&l_t$&$q_t^{(1)}$&$n\!p_t^{(1)}$&$p\&l_t^{(1)}$& $cp\&l_t^{(1)}$&$q_t^{(2)}$&$n\!p_t^{(2)}$
&$p\&l_t^{(2)}$&$cp\&l_t^{(2)}$\\ \hline\hline
                                   &\$100 & 8  & 8   & -       & 7   & 7                 & -   & -                    &  1  & 1  							& -  & -\\
                                   &\$130 & 2  &10   & \$240   & 2   & 9                 & 210 & \$210                &  0  & 1  							& 30 & \$30\\
\raisebox{1.25ex}[0pt]{\nth{1} day}&\$150 & -4 & 6   & \$440   & -4  & 5                 & 180 & \$390                &  0  & 1  							& 20 & \$50\\
                                   &\$140 & -10& -4  & \$380   &-9  & -4                & -50 & \$340                & -1  & 0  							&-10 & \$40\\ \hline\hline
                                   &\$110 & -4 & -8  & \$500   & -3  & -7                & 120 & \$460                & -1  & -1  							& 0  & \$40\\
                                   &\$115 & -4 & -12 & \$460   & -4  & -11               & -35 & \$425                & 0   & -1  							& -5 & \$35\\
\raisebox{1.25ex}[0pt]{\nth{2} day}&\$110 & -4 & -16 & \$520   & -3  & -14               &  55 & \$480                & -1  & -2  							& 5  & \$40\\
                                   &\$80  & 16 & 0   & \$1000  & 14  & \MyColorBoxO{0}   & 420 & \MyColorBoxY{\$900}  &  2  & \MyColorBoxO{0}& 60  & \MyColorBoxY{\$100}\\
\hline
\end{tabular}
\caption{Allocation utilizing \texttt{FOUR}}\label{table:four}
\end{table}
As shown in Table \ref{table:four}, allocations yield a perfect P\&L for each account and both accounts are flat at the end of \nth{2} day. The methodology is fair and equitable. Important to note that there is no need for any manual intervention  or trial and error on rounding. The algorithm performs optimal rounding for each bunched filled order (buy or sell) by minimizing return variations across accounts through the \texttt{ML} algorithm. 

\section{Assessment based on our numerical results}

In Table \ref{table:summary}, we show summary of trade allocation into two accounts using rounding, \texttt{HPHA}, \texttt{APS} and \texttt{FOUR}. Recall in a perfect scenario, after allocating bunched orders into the two accounts, the net position of both accounts should be zero, with realized P\&L \$900 and \$100 for larger account and smaller one accounts (\$1000 for bunched orders). We have highlighted cases where accounts P\&Ls are not fair and equitable or the net positions of the accounts are not consistent with the bunched orders, i.e. one account is long versus the other being short whereas bunched orders are flat.

\begin{table}[H]
\begin{center}
\begin{tabular}{|c|c||c|c||c|c||c|c|} \hline
   \multicolumn{2}{|c||}{\raisebox{-1.25ex}[0pt]{Method}}  & \multicolumn{2}{c||}{Bunched Orders} &   \multicolumn{2}{c||}{\nth{1} account}   &     \multicolumn{2}{c|}{\nth{2} account}   \\ \cline{3-8}
   \multicolumn{2}{|c||}{       } &  $n\!p_t$ & $cp\&l_t$      & $n\!p^{(1)}_t$&$cp\&l^{(1)}_t$ & $n\!p^{(2)}_t$ &$cp\&l^{(2)}_t$  \\ \hline\hline
                                        &   simple             &  0  &  \$1000    &  \MyColorBoxO{-2}  &  \MyColorBoxY{\$960}   &  \MyColorBoxO{2}    &   \MyColorBoxY{\$40}    \\
rounding                                &   \nth{1} trial on rounding &  0  &  \$1000    &  0   &  \MyColorBoxY{\$860}   &  0    &   \MyColorBoxY{\$140}   \\
                                        &   \nth{2} trial on rounding &  0  &  \$1000    &  0   &  \$900   &  0    &   \$100   \\ \hline\hline
\raisebox{-1.25ex}[0pt]{\texttt{HPHA}}  &      no adjustment          &  0  &  \$1000    &  \MyColorBoxO{1}   &  \MyColorBoxY{\$1010}  &  \MyColorBoxO{-1}   &   \MyColorBoxY{-\$10}   \\
                                        &   non-trivial adjustment    &  0  &  \$1000    &  0   &  \MyColorBoxY{\$870}   &  0    &   \MyColorBoxY{\$130}   \\ \hline\hline
\raisebox{-1.25ex}[0pt]{\texttt{APS}}   &     no adjustment           &  0  &  \$1000    &  \MyColorBoxO{1}   &  \MyColorBoxY{\$1011.5}&  \MyColorBoxO{-1}   &   \MyColorBoxY{-\$11.5} \\
                                        &   non-trivial adjustment    &  0  &  \$1000    &  0   &  \MyColorBoxY{\$874.3} &  0    &   \MyColorBoxY{\$125.7} \\ \hline\hline
\texttt{\textcolor{blue}{FOUR}}         &\texttt{ML}-based allocation &  0  &  \$1000    &  0   &  \$900   &  0    &   \$100   \\ \hline
\end{tabular}
\caption{Summary of trade allocation into two accounts using rounding, \texttt{HPHA}, \texttt{APS} and \texttt{FOUR}}\label{table:summary}
\end{center}
\end{table}

Overall, we see for our sample portfolio we cannot obtain fair and equitable treatment for the two accounts unless we make some adjustment to rounding. Adjustments should not be based on trial and error as it can become intractable, very time-consuming, and not feasible as the number of accounts increases and trades sizes are varying at different prices. Moreover, it cannot guarantee the obtained adjustment based on trial and error is optimal.

We think practitioners should ask some baseline questions about the allocation methodology used for allocation of bunched orders or \texttt{block trades} into multiple accounts: (a) Does the method yield similar performance among all accounts? (b) Are the results from the method agnostic to the account size, number of accounts, market conditions and market volatility? Irrespective of number of accounts, size of the accounts, and market conditions, the allocation method should generate fair and equitable results. (c) Does the allocation method require any manual interventions or adjustments to make allocation fair and equitable across all accounts? Where, it seems many of the prevalent methods do not give a clean answer. 
 
The objective in \texttt{FOUR} is based on finding the most optimal or post possible adjustment to rounding for each bunched order by minimizing variation across all accounts which is based on an \texttt{ML} algorithm and does not need any manual intervention or adjustment.

\section{Conclusion}\label{sec:con}
Investors choose vehicles given their preferences but expect the returns to be aligned across the options.  For example, an investor may choose an \texttt{SMAs} to potentially exert a greater degree of control over their investment but expect it to track the gross performance of the underlying \texttt{Fund} or other comparable investments.  However, unlike investing in a \texttt{Fund}, where all investors receive identical gross returns, any \texttt{SMA} is subject to trade allocation risk, and its return could diverge from that of the \texttt{Fund}, other \texttt{SMAs} or comparable options (as in positively or negatively for the participating accounts as the bunched orders are allocated using a particular methodology).  This problem is only going to exacerbate given the increasing popularity of direct indexing and such customizable account strategies.  The reason for this is that trades need to be allocated to each account before the calculation of returns, and this step may involve rounding, which leads to biased allocation of shares and returns as illustrated through aforementioned examples. This allocation/return bias is directly proportional to the number and profitability of trades, and inversely proportional to average trading size.  As a result, without frequent correction, the bias, and return disparity increases with increasing number of trades. In particular because of rounding, smaller trades and remainders will be allocated to the largest accounts, which results in biasing allocation and returns against smaller accounts. 

To remove these biases, we have presented the ``{\em fair optimal unbiased rounding}'' (\texttt{FOUR}), an \texttt{ML} algorithm, for allocation which dynamically change the rounding bias for each trade to more evenly distribute the returns across all accounts over time. In effect, \texttt{FOUR} keeps track and corrects for under/over allocation of returns over time. \texttt{FOUR} leverages advancements in financial engineering as a robust allocation method that (a) supports fair and equitable allocation across all accounts, (b) removes biases to account size, volume, clients, other, (c) allows automated allocations that are easily auditable, and (d) enables fast reconciliations with no manual interventions or manipulations.

\section*{Acknowledgments}

We are very grateful to Satyan Malhotra, CEO of \texttt{ASK}{\scriptsize{2}}.ai and Marco Ricciardulli for their thorough review of this paper and very useful comments and suggestions. Errors are our own responsibility.

\appendix

\section*{Appendix}

We have extensively tested \texttt{FOUR} against the cited methodologies with \& without rotation for both real \& simulated portfolios with various trade sizes and number of accounts. For all these tests, \texttt{FOUR} yielded fair \& equitable returns among all accounts. On the other hand, the cited methodologies (without human interventions and trial \& error) performed very poorly with large discrepancies among accounts in all scenarios. In all of our tests, we have observed discrepancies exacerbated as we increase the number of multiple accounts. For any inquiry regarding testing \texttt{FOUR} for your sample portfolio or need more information on our testing results, email info@ask2.ai.

\bibliographystyle{plain}
\bibliography{bibliography}

\end{document}